\documentclass[nofootinbib, 11pt, onecolumn, superscriptaddress, notitlepage, tightenlines, aps]{revtex4-1} 
\usepackage{stmaryrd}
\usepackage{tabstackengine}
\stackMath
\makeatletter
\renewcommand\TAB@delim[1]{\scriptstyle#1}
\makeatother
\setstackgap{S}{2pt}
\usepackage[sort&compress]{natbib}
\usepackage[utf8]{inputenc}
\usepackage[english]{babel}
\usepackage[mathscr]{euscript} 

\DeclareFontFamily{OT1}{pzc}{}
\DeclareFontShape{OT1}{pzc}{m}{it}{<-> s * [1.10] pzcmi7t}{}
\DeclareMathAlphabet{\mathpzc}{OT1}{pzc}{m}{it}
\usepackage{adjustbox}
\usepackage{lipsum}
\usepackage[sc]{mathpazo}
\usepackage[T1]{fontenc}
\usepackage{amsmath}
\usepackage{relsize}
\usepackage{bigints}
\usepackage[usenames, dvipsnames]{xcolor}
\definecolor{Indigo}{RGB}{70,50,120}
\usepackage{hyperref}
\hypersetup{urlcolor=Indigo, colorlinks=true,citecolor=Indigo,linkcolor=Indigo} 
\usepackage{graphicx}
\usepackage{dcolumn}
\usepackage{bm}
\usepackage{amssymb}
\usepackage{comment}
\usepackage{amsthm}
\usepackage{booktabs}
\usepackage{epstopdf}
\usepackage{float}
\usepackage{tikz, pgfplots}
\usetikzlibrary{matrix}
\usepgfplotslibrary{fillbetween}
\usetikzlibrary{calc, automata, chains, arrows.meta, quotes}
\usepackage{enumitem}
\usepackage[linesnumbered,ruled,vlined]{algorithm2e}

\usepackage{thmtools}
\usepackage{thm-restate}

\usepackage[capitalise]{cleveref} 

\newcommand{\be}{\begin{eqnarray}}
\newcommand{\ee}{\end{eqnarray} }
\newcommand{\benn}{\begin{eqnarray*}}
\newcommand{\eenn}{\end{eqnarray*}}
\newcommand{\txt}{\textrm}
\newcommand{\X}{\rangle\!\langle}

\newcommand{\D}{\txt{d}} 
\newcommand{\bpr}{\begin{proof}}
\newcommand{\epr}{\end{proof}}

\newtheorem*{theorem*}{Theorem}
\newtheorem*{corollary*}{Corollary}

\newtheorem{definition}{Definition}
\newtheorem{definition*}{Definition}[section]

\newtheorem{lemma}{Lemma}

\newtheorem*{prop*}{Proposition}
\DeclareMathOperator{\Tr}{tr}
\DeclareMathOperator{\Comm}{Comm}

\SetCommentSty{mycommfont}

\definecolor{Crimson}{RGB}{220, 20, 60}
\definecolor{pinkred}{RGB}{180, 10, 100}
\definecolor{Blue1}{RGB}{100, 20, 200}
\definecolor{Blue2}{RGB}{100, 20, 100}

\begin{document}
\title{Haar-random and pretty good measurements for Bayesian state estimation}
\author{Maria Quadeer\footnote{mariaquadeer@gmail.com}}
\affiliation{School of Physical \& Mathematical Sciences, Nanyang Technological University, Singapore}
\date{\today}
\begin{abstract}
We study Haar-random bases and pretty good measurement for Bayesian state estimation. Given $N$ Haar-random bases we derive a bound on fidelity averaged over IID sequences of such random measurements for a uniform ensemble of pure states. For ensembles of mixed qubit states, we find that measurements defined through unitary 2-designs closely approximate those defined via Haar random unitaries while the Pauli group only gives a weak lower bound. For a single-shot-update, we show using the Petz recovery map for pretty good measurement that it can give pretty good Bayesian mean estimates.
\end{abstract}

\maketitle 


\section{Introduction}
The problem of quantum state estimation is an estimation theory formulation of the related problem of quantum state tomography \cite{helstrom_1969}. Quantum state tomography deals with the reconstruction of the full density matrix that describes a physical system which typically requires a large number of copies of the system. In contrast quantum state estimation allows us to define functions called \emph{estimators} on a single copy and frees us from the constraint of requiring many copies of the system. The state estimation formalism is a single-shot framework that allows us to ask questions about optimality of estimators as well as measurements under a chosen measure of distance between the estimate and the true state \cite{Lehmman98}.

Optimality of an estimator depends on the chosen measure of distance, and one can define different criteria of optimality. For instance, the Bayesian mean estimator minimizes the average risk for Bregman divergences \cite{Banerjee} whereas the corresponding optimal estimator for fidelity is still unknown \cite{Keung-Ferrie_2015}. Additionally, there is the problem of finding optimal measurements given an optimal estimator which is generally hard except when the states being estimated belong to an orbit generated by a unitary group \cite{Quadeer2019minimaxquantumstate}. In that case the problem simplifies to that of covariant state estimation \cite{holevo82}, where it has been shown that covariant measurements are minimax, i.e. they minimise the minimax risk \cite[Theorem 4.3]{Quadeer2019minimaxquantumstate}. Given the complexity around proving such results we can ask: are there specific measurement choices that give reasonable estimates for arbitrary quantum states?

One such choice is that of Haar-random measurements that arise in the context of distinguishing states of an ensemble. The question is if there exists a single positive operator valued measure (POVM) that gives reasonably large total variation distance between every pair of states in the ensemble. It was shown by P. Sen \cite{sen_random_2006} that measuring any two low-rank quantum states, in a Haar-random orthonormal basis, typically yields probability distributions that are separated by at least a universal constant times the Frobenius distance between the states with a high probability of success. Given that Haar-random orthonormal bases are not efficiently implementable \cite{pennylane_tut}, this led to the question of finding pseudo-random orthonormal bases. 

The conception of unitary $t$-designs \cite{unitary-t-designs-GAE_2007, unitary-2-design_Scott_2008, unitary-t-designs_Roy-Scott_2009,bannai2018unitary} followed around the same time and a derandomization of Sen's result was given \cite{derandomize_ambainis-emerson_2007}. Unitary $t$-designs are finite sets of unitary matrices that mimic the properties of the Haar measure on the unitary group in that a uniform weight over such a set gives the same average as the Haar measure for polynomials of degree $\leq t$. In this sense a unitary $t$-design is a pseudo-random quantity. Maximal sets of Mutually Unbiased Bases (MUBs) that form a unitary 2-design are optimal for state tomography of large but finite ensembles of states under \emph{information gain} \cite{Wootters-Fields_MUBs_2006}. While a construction for a maximal set of $d+1$ MUBs is known in prime power dimensions \cite{Wootters-Fields_MUBs_2006}, whether such a maximal set exists for non-prime power dimensions is still an open question \cite{grassl_sic-povms_2009, brierley_maximal_2008}. 

Haar random measurements also arise in the context of classical shadow tomography where one is interested in estimating expectation values of observables up to a given accuracy \cite{paini2019approximate, predicting-many-propr-few-measurements_Huang-et-al_2020}. Classical shadow tomography is set-up such that one only requires finite number of such random measurements and derandomization is not a point of concern in such frameworks. It was shown by Huang et. al. \cite{predicting-many-propr-few-measurements_Huang-et-al_2020} that by obtaining an approximate classical description of the state using $\log M$ random measurements one is able to accurately predict $M$ linear functions of the state with a high success probability. 

Another choice of measurement given an ensemble of quantum states is the pretty good measurement, the POVM elements of which can be expressed in terms of the states of the ensemble itself \cite{watrous_2018}. Pretty Good Measurement (PGM) arises in the general context of distinguishing states of an ensemble. Given an unknown state one performs the PGM and declares the state to be the one corresponding to the measurement outcome \cite{Petz-recovery_PGM_Barnum-Knill_2002, sen_random_2006, PGM_Harrow-Winter_IEEE_2012, Hausladen-Wootters_PGM_1994}. While it is optimal for distinguishing orthogonal states of an ensemble it is only near-optimal for arbitrary ensembles in the sense that PGM incurs an error in correct prediction that is at most twice as large as the optimal error in prediction \cite[Theorem 3.10]{watrous_2018}\footnote{The earlier proofs \cite{Hausladen-Wootters_PGM_1994, Petz-recovery_PGM_Barnum-Knill_2002} required some constraints on the ensemble. A general proof was given in Watrous' book \cite[Theorem 3.10]{watrous_2018}}.


In this work we study state tomography in the estimation theory framework. Given a POVM $\mathcal{M}$, we define an estimator of a quantum state as a function $\hat{\rho}: X \mapsto \mathcal{D}(\mathcal{H})$, where $X$ represents the alphabet of measurement outcomes and $\mathcal{D}(\mathcal{H})$ denotes the set of density matrices. We consider \emph{single-copy} (unentangled) measurements where one measures quantum systems sequentially. We assume that a process exists for preparing independent and identically distributed (IID) copies of quantum systems in a state $\rho$, which are discarded post-measurement. The sample complexity is therefore the same as the number of measurements performed. We choose a distance measure to denote the error in the estimate $\hat{\rho}$ of the unknown state $\rho$ chosen from a prior ensemble $\mathcal{E}\sim\{p_i, \rho_i\}$, and define risk as the expected error with respect to the measurement outcomes. As finding optimal measurements for arbitrary states remains an open problem, we consider (sequences of) Haar-random measurement bases and PGM to see if these can give reasonable estimates for arbitrary ensembles of states. In particular, for the case of Haar-random measurements, we ask: is it possible to derandomize the process, for example, can Haar-random measurements be replaced by a unitary $t$-design? For the case of PGM, since we know that it is near-optimal in terms of the probability of correctly distinguishing states of an ensemble, the question is: can Bayesian mean estimation be pretty good under PGM in an appropriately defined sense?


We measure error in terms of fidelity and use the Bayesian mean estimator \cite{Blume-Kohout2010} since \textit{a}. it has a closed form expression and is the mean with respect to the posterior distribution given a prior distribution over the set of density matrices $\mathcal{D}(\mathcal{H})$, \textit{b}. it fits an algorithmic framework due to the Bayes update rule, and is independent of the measurement procedure in contrast to other algorithms \cite{brandao_fast_2021} where the convergence of the algorithm depends on the measurement procedure satisfying some generic properties, and \textit{c}. convergence of the posterior distribution is guaranteed under sufficient regularity conditions on the prior \cite{convergence-rates-posterior-Bayes-rule_Ghoshal-et.al_2000}. 

With the above set-up, we first consider Haar-random measurements by defining an orthonormal random basis vis-à-vis the columns of a Haar-random unitary matrix, which can be generated numerically \cite{mezzadri_how_2007}. We obtain the estimates from these random orthonormal bases for ensembles of pure and mixed states under a Bayesian updates algorithm. Given $N$ copies of a system in the state $\rho$, we update the prior after each measurement to obtain the posterior. At the $N$-th step we evaluate the Bayes estimator which is the mean with respect to the posterior. For ensembles of mixed states we find that the expression for average risk involves Haar averaging of a rational function of the unitaries. This implies that it is impossible to find a $t$-design that would derandomize the Bayesian update procedure. However when restricted to pure states the expression simplifies and we are able to obtain an upper bound for average risk (in terms of infidelity). We compare measurements defined through the Pauli group (1-design), a unitary 2-design, the Clifford group unitaries (unitary 3-design), and Haar random bases for the case of a qubit. We find that unitary 2-designs closely approximate Haar random unitaries under average fidelity for ensembles of mixed qubit states while the Pauli group (that forms a 1-design) only gives a weak lower bound.

Next, we consider pretty good measurement (PGM), which can always be constructed using the states of an ensemble, with each POVM element corresponding to a specific state. As discussed earlier, PGM is pretty good for distinguishing states of an ensemble. The strategy employed to distinguish  states using PGM assigns outcomes to the corresponding states but we know that this is quite restrictive and indeed does not work for specific ensembles \cite{Hausladen-Wootters_PGM_1994}. Instead of asking for optimal measurements for distinguishing states of an ensemble, we ask if Bayesian mean estimates can be pretty good under PGM. Thus instead of identifying states with outcomes we give Bayesian mean estimates corresponding to the outcomes. This intuition is rooted in the fact that the Petz recovery map corresponding to a preparation channel $\mathcal{A}$ \cite{Petz-recovery_PGM_Barnum-Knill_2002}, that maps classical registers $a$ to the corresponding states $\rho_a$ of the ensemble, outputs a mean of the classical states with respect to the posterior distribution corresponding to PGM. We show this by first obtaining the covariance of the output distribution for PGM under a Bayesian update.

The remainder of this paper is structured into four sections. Section II provides essential notation and definitions. Section III presents theoretical results and discusses the numerical implementation of the Bayesian updates procedure with Haar random measurements. Section IV explores the application of PGM to obtain pretty good single-shot Bayesian mean estimates. Lastly, Section V summarizes the work and outlines potential avenues for future research. The numerical codes used to generate data shown in this work is available on GitHub \cite{Quadeer_BME-RMB-PGM}.

\section{Formalism}\label{sec:formalism}

We consider finite $d$-dimensional quantum systems and denote the corresponding Hilbert spaces by $\mathcal{H}$. The states of the system are described by density matrices that are positive semi-definite matrices with trace one. We denote the set of positive semidefinite matrices by $\mathcal{P}(\mathcal{H})$ and the set of density matrices by $\mathcal{D}(\mathcal{H}) \subset \mathcal{P}(\mathcal{H})$. The set of pure states i.e. normalized vectors in $\mathcal{H}$ on the other hand is denoted by $\mathcal{S}(\mathcal{H})$. We will refer to either of these depending on the context.

\begin{definition}[Ensemble]
	 An ensemble $\mathcal{E}$ of quantum states is defined on an alphabet $\Gamma$ through
	 \be\label{eq:ensemble}
	 	\tilde{\rho}: \Gamma \mapsto \mathcal{P}(\mathcal{H}), ~~~ \Tr \Bigg(\sum_{a \in \Gamma}\tilde{\rho}(a)\Bigg) = 1,
	 \ee
	 where $\tilde{\rho}(a)$ denotes a quantum state $\rho_a$ together with a probability $p_a$, 
	\be 
		\rho_a =\frac{\tilde{\rho}(a)}{\Tr(\tilde{\rho}(a))}, ~~~p_a = \Tr(\tilde{\rho}(a)).
	\ee 
\end{definition}
\begin{definition}[Measurement]
	A measurement $\mathcal{M}$ is defined on an alphabet $X$ as\footnote{We may also refer to X as a random variable taking values in X which represents the measurement outcomes.}
	\be 
		M: X \mapsto \mathcal{P}(\mathcal{H}), ~~~ \sum_{x \in X} M(x) = \mathbb{1}.
	\ee 
\end{definition}
Measurement in the standard computational basis is defined through the computational basis as the set $\{|x\X x|\}_{x\in X}$. An estimator of a quantum state is a function $\hat{\rho}: X \mapsto \mathcal{D}(\mathcal{H})$. Given an unknown state $\rho$ drawn from the ensemble $\mathcal{E}$, the error in its estimate $\hat{\rho}(x)$ is measured in terms of a distance function $D: \mathcal{D(\mathcal{H})} \times \mathcal{D(\mathcal{H})} \mapsto \mathbb{R_+}$. For the rest of this paper we will assume that all such functions are convex in the second argument. The risk of the estimator is defined as the expected value of the distance,
\be\label{eq:risk}
R(\rho, \hat{\rho}) = \sum_{x\in X} p(x|\rho) D\big(\rho,\hat{\rho}(x)\big),
\ee
and the average risk of the estimator is defined as the mean risk with respect to a prior distribution $\bold{p}$ for an ensemble of states,
\be \label{eq:average-risk}
r(\bold{p}, \hat{\rho}) = \sum_a p_a ~R(\rho_a, \hat{\rho}).
\ee 
Upon measurement of a single copy of the unknown state, let us say outcome $x$ was observed. A posterior distribution over states $\bold{p_{A|X}}$ can then be obtained via the Bayes rule,
		\be \label{eq:Bayes-rule}
			p(a|x)=\frac{p(x|a)p_a}{p(x)},~~~~p(x) = \sum_{a} p(x|a)p_a,
		\ee 
		where the conditional distribution $\bold{p_{X|A}}$ is given by the Born's rule, 
		\be 
			p(x|a) = \Tr(M(x)\rho_a).
		\ee 
We will be working with the Bayesian mean estimator which is the mean with respect to the posterior distribution \cite{Blume-Kohout2010},
		\be \label{eq:Bayes-est}
			\hat{\rho}_B (x) = \sum_a p(a|x)\rho_a.
		\ee 	
The set of unitary matrices for a given $d$ is denoted by $\mathbb{U}(d)$, and we denote the set of all linear operators on $\mathcal{H}$ by $L(\mathcal{H})$. 
\begin{definition}[Commutant]
The commutant of a subset of operators $\mathcal{A} \subseteq L(\mathcal{H})$ is the set 
	\be 
		\Comm\{A\} = \{Y \in L(\mathcal{H}) : [X, Y] = 0 ~\forall X \in \mathcal{A}\}.
	\ee 	
\end{definition}
The symmetric subspace of the composite space $\mathcal{H}^{\otimes N}$ is denoted by $\mathcal{H}^{\varovee N}$, the orthogonal projector onto which is denoted by $\Pi_+^{d, ~N}$. The dimension of the symmetric subspace  is $\binom{N+d-1}{N}$.
\section{Bayesian mean estimation - Haar-random measurements}\label{sec:BME-RMB}
Let us assume that we are given a system in an unknown quantum state $\rho$, chosen uniformly at random from an ensemble $\mathcal{E}$, and that we want to obtain an estimate for this state using Bayesian mean estimation by performing single copy measurements in Haar-random bases. A Haar-random unitary $\mathcal{U}$ defines a Haar-random basis via its columns, $\{\mathcal{U}|x\rangle\langle x|\mathcal{U}^\dagger\}_{x\in X}$. Thus, the conditional probability distribution is
 \be\label{eq:p(x|i)}
 p_{\mathcal{U}}(x|\rho) =\Tr\big(\mathcal{U}|x\rangle\langle x|\mathcal{U}^\dagger \rho\big).
 \ee 
As we are looking at Haar-random unitary matrices, the quantity of interest would be the Haar average of the risk over the
unitary group,
\be \label{eq:Haar-av-risk}
\int\D\mathcal{U} R_{\mathcal{U}}\big(\rho, \hat{\rho}_B\big).
\ee 
However, this corresponds to the Haar average of the point-wise (single-copy outcome) Bayes estimator, and as we are interested in a sequence of updates we need the Haar average of the risk for the Bayesian estimator with $N$ updates. Let us consider \cref{fig:algo1}. Average risk as defined in the algorithm is the average of risk over the given ensemble $\mathcal{E}$ with respect to the posterior distribution, computed as a function of an IID sequence $\mathcal{N}$ of Haar-random measurement bases. The risk of the Bayes estimator for a sequence $\mathcal{N}$ of measurements can be written as 
\be \label{eq:Bayes-est-vector}
R_{\mathcal{N}}\big(\rho, \hat{\rho}_B\big) = \sum_{x}p_{\mathcal{U}}(x|\rho)D\big(\rho, \hat{\rho}^{\mathcal{N}}_B(x)\big),
\ee 
where the conditional distribution corresponds to the last measurement basis defined through $\mathcal{U}$, given by \eqref{eq:p(x|i)}, and the Bayes estimator is given by
\be\label{eq:Bayes-est-N-outcomes}
\hat{\rho}^{\mathcal{N}}_B(x) = \sum_a p^{\mathcal{N}}(a|x)\rho_a,
\ee
where the posterior distribution, obtained via the Bayes update rule for $\mathcal{N}$, is evaluated by updating the prior given measurement outcomes $x_1\cdots x_{N-1}$,
\be \label{eq:posterior-N-outcomes}
p^{\mathcal{N}}(a|x)=\frac{p_{\mathcal{U}}(x|a)p^{\mathcal{U}_{N-1}\cdots\mathcal{U}_1}_a}{p_{\mathcal{U}}(x)} = \frac{p_{\mathcal{U}}(x|a)p_{\mathcal{U}_1}(x_1|a)\cdots p_{\mathcal{U}_{N-1}}(x_{N-1}|a)p_a}{p_{\mathcal{U}}(x)p_{\mathcal{U}_1}(x_1)\cdots p_{\mathcal{U}_{N-1}}(x_{N-1})}.
\ee
The denominator above is given by the normalization constant,
\be \label{eq:denominator}
 \prod_{i=1}^N p_{\mathcal{U}_i}(x_i) = \Tr \Bigg(\bigotimes_i^N \mathcal{U}_i|x_i\X x_i|\mathcal{U}_i^\dagger \sum_b p_b \rho_b^{\otimes N} \Bigg).
\ee 
We consider the Haar average of risk as in Eqn. \eqref{eq:Bayes-est-vector} over the sequence of measurements $\mathcal{N}$ for \cref{fig:algo1},
\be \label{eq:av-risk}
\int\D\mathcal{N} R_{\mathcal{N}}\big(\rho, \hat{\rho}^{\mathcal{N}}_B\big)&=& \int\D\mathcal{N} \sum_{x}~p_{\mathcal{U}}(x|\rho)D\big(\rho, \hat{\rho}^{\mathcal{N}}_B(x) \big).
\ee 
The expression for the Bayes estimator using \eqref{eq:Bayes-est-N-outcomes} and the convexity of $D$ in the second argument implies the inequality,
\be \label{eq:Haar-av-risk-n-design-3}
\int\D\mathcal{N} R_{\mathcal{N}}\big(\rho, \hat{\rho}^{\mathcal{N}}_B\big) &\leq &   \int \D \mathcal{N}  \sum_{x,~a}~  D\big(\rho, \rho_a\big)p_{\mathcal{U}}(x|\rho) p^{\mathcal{N}}(a|x).
\ee
The fact that the sequence $\mathcal{N}$ is IID and the expression for the posteriors, i.e. Eqns. \eqref{eq:p(x|i)} and \eqref{eq:posterior-N-outcomes} together imply
\be \label{eq:Haar-av-risk-n-design-4}
\int\D\mathcal{N}~ R_{\mathcal{N}}\big(\rho, \hat{\rho}^{\mathcal{N}}_B\big) &\leq & 
\mathlarger{\mathlarger{\mathlarger{\sum}}}_{x,~a}~p_a ~D\big(\rho, \rho_a\big) \bigintsss \D\mathcal{U}\D\mathcal{U}_{N-1}\cdots \D\mathcal{U}_1 \Tr \Bigg(~\mathcal{U}^{\otimes 2} \big(|x\X x|\big)^{\otimes 2}\big(\mathcal{U}^\dagger)^{\otimes 2} (\rho \otimes \rho_a)\Bigg) \times \notag \\
& & ~~~~~~~~~~~~~~~~~~~~~~~~~~~~~~~~\frac{\prod_{j=1}^{N-1}\Tr\big(\mathcal{U}_j|x_j\rangle\langle x_j|\mathcal{U}_j^\dagger \rho_a\big)}{\prod_{i=1}^N p_{\mathcal{U}_i}(x_i)}.
\ee 
In general the integral in \eqref{eq:Haar-av-risk-n-design-4} involves Haar averaging of a fractional polynomial in degrees of $\mathcal{U}$ and $\mathcal{U}^\dagger$, and it is not known how to do such integrals. However, when estimating Haar distributed pure states, one can circumvent this issue and obtain an upper bound. We state and prove this in the form of \cref{lem:pure}.

\begin{algorithm}[h!]
\SetKwInput{KwInput}{Given}                           
\SetKwInput{KwOutput}{Parameters}                     
\DontPrintSemicolon
  
  \KwInput{$\mathcal{N}$, $\mathcal{E}$}
  \KwOutput{sample complexity $|\mathcal{N}|=N$}
  \KwData{average risk}\;

  \SetKwFunction{FMain}{Main}
  \SetKwFunction{FBayes}{Bayes Update}
  \SetKwFunction{FHaar}{Haar-random $\mathcal{U}$}
  \SetKwFunction{FInvSample}{Inverse Transform Sampling}
  
  \SetKwProg{Fn}{func}{:}{}
  \Fn{\FHaar{}}{
        \KwRet basis\;
  }
  \;
  
  \SetKwProg{Fn}{func}{:}{}
  \Fn{\FInvSample{\textbf{p}}}{
        \KwRet x\;
  }
  \;
  
  \SetKwProg{Fn}{func}{:}{\KwRet}
  \Fn{\FMain{}}{
  		
  		\tcc{\color{Indigo}{perform Bayesian updates for $\mathcal{N}$ after creating ensemble of states $\mathcal{E}$}}   
  		\For{(n = 0, n < N)}{
  			basis = \FHaar{}\;
  			\tcc{\color{Indigo}{evaluate conditional $\bold{p_{X|A}}$  and total probabilities $\bold{p_X}$ and measure}}
  			outcome = \FInvSample{$\bold{p_X}$}\;
  			\tcc{\color{Indigo}{update prior given outcome}}
  			}
  			
  			\tcc{\color{Indigo}{evaluate Bayes estimator and average risk}}
  			
        \KwRet 0 \;
  }
\caption{Bayesian mean estimation with Haar-random bases\footnote{Note that we inverse sample the measurement outcomes using the total distribution and not the conditional distribution since we are interested in how the algorithm behaves broadly under the specific choice of measurement considered here. This allows us to avoid making a choice for the input state and further analysing the convergence as a function of this choice.}} 

\label{fig:algo1} 

\end{algorithm}


\begin{lemma}[Bayesian mean estimation with $N$ random measurement bases for pure states under fidelity]\label{lem:pure} 
Given a sequence of $N$ Haar-random orthonormal measurement bases and an ensemble of Haar distributed pure states, the average risk for \cref{fig:algo1} expressed in terms of infidelity satisfies the inequality 
\be \label{eq:lem1}
\int\D\mathcal{N} R_{\mathcal{N}}\big(|\psi \X \psi |, \hat{\rho}^{\mathcal{N}}_B\big) &\leq & 1 - \mathrm{f}(d,~N),
\ee 
where $\mathrm{f}(d,~N) = \frac{(d+3)(N+d-1)!}{d^{N} (d+1)^2 N! (d-1)!}$.
\end{lemma}
\bpr
The risk of the Bayes estimator for $\rho = |\psi \X \psi |$, chosen from an ensemble of Haar distributed pure states, expressed in terms of infidelity is linear. Thus the inequality in \eqref{eq:Haar-av-risk-n-design-3} can be replaced by an equality and we have
\be \label{eq:lem1:1}
\int\D\mathcal{N} R_{\mathcal{N}}\big(|\psi \X \psi |, \hat{\rho}_B\big) &=& 1 - \int \D \mathcal{N} \mathlarger{\mathlarger{\sum}}_{x,~a} ~p_{\mathcal{U}}(x||\psi \X \psi |)p^{\mathcal{N}}(a|x)\langle \psi | \rho_a | \psi \rangle.
\ee 
We consider Eqn. \eqref{eq:Haar-av-risk-n-design-4} and note that $\sum_b p_b \rho_b^{\otimes N} \in \Comm \{U^{\otimes N} | U \in \mathbb{U}(d)\}$ as we have an ensemble of Haar distributed pure states. This implies \cite[Lemma 7.24]{watrous_2018} that
\be \label{eq:lem1:2}
\sum_b p_b \rho_b^{\otimes N} = \frac{\Pi_+^{d, ~N}}{\mathrm{dim}(\mathcal{H}^{\varovee N})},
\ee 
and we use this fact to obtain the upper bound by plugging it into \eqref{eq:denominator} to obtain
\be \label{eq:lem1:3}
\prod_{i=1}^N p_{\mathcal{U}_i}(x_i) 
&=& \frac{1}{\mathrm{dim}(\mathcal{H}^{\varovee N})}\Tr \Bigg(\bigotimes_{i=1}^N \mathcal{U}_i |x_i\X x_i| \mathcal{U}_i^\dagger ~\Pi_+^{d, ~N}\Bigg).
\ee 
Now as $\Pi_+^{d, ~N}$ is an orthogonal projector onto the symmetric subspace of $\mathcal{H}^{\otimes N}$, we have 
\be \label{eq:lem1:4}
\prod_{i=1}^N p_{\mathcal{U}_i}(x_i) 
&\leq& \frac{1}{\mathrm{dim}(\mathcal{H}^{\varovee N})}\Tr \Bigg(\bigotimes_{i=1}^N \mathcal{U}_i |x_i\X x_i| \mathcal{U}_i^\dagger\Bigg) \notag \\
&=& \frac{1}{\mathrm{dim}(\mathcal{H}^{\varovee N})}.
\ee 
We thus obtain an upper bound to Eqn. \eqref{eq:lem1:1} which is a polynomial integrand,
\be \label{eq:lem1:4}
\int\D\mathcal{N} R_{\mathcal{N}}\big(\rho, \hat{\rho}^{\mathcal{N}}_B\big) &\leq & 1 - \frac{\mathrm{dim}(\mathcal{H}^{\varovee N})}{d^{N-1}}~\mathlarger{\mathlarger{\sum}}_{x,~a} ~ p_a \langle \psi | \rho_a | \psi \rangle \times \notag \\ 
&~& ~~~~~~~~~~~~~~~~~~~~~~~~ \int \D \mathcal{U} \Tr \big( \mathcal{U}^{\otimes 2}|x\X x|^{\otimes 2} (\mathcal{U}^\dagger)^{\otimes 2} |\psi \X \psi| \otimes \rho_a \big),
\ee 
as the remaining $N-1$ integrals $\int \D \mathcal{U}_j~ \mathcal{U}_j|x_j\rangle\langle x_j|\mathcal{U}_j^\dagger$ lie in the $\Comm\{\mathbb{U}(d)\}$ and thus equal $\frac{\mathbb{1}}{d}$ by Schur's lemma \cite[pg. 152]{weyl1950theory}. We can now evaluate the integral using the Schur-Weyl duality \cite[Proposition A.3]{LN22} for the Haar average,
\be 
\int\D\mathcal{N} R_{\mathcal{N}}\big(\rho, \hat{\rho}^{\mathcal{N}}_B\big) &\leq & 1 - \frac{\mathrm{dim}(\mathcal{H}^{\varovee N})}{d^{N-1}}~\mathlarger{\mathlarger{\sum}}_{x,~a} ~ p_a \langle \psi | \rho_a | \psi \rangle \Tr \Big( \big(\alpha \mathbb{1} + \beta \mathbb{W}\big) |\psi \X \psi| \otimes \rho_a \Big) \notag \\
&=& 1 - \frac{\mathrm{dim}(\mathcal{H}^{\varovee N})}{d^{N-2}}\mathlarger{\mathlarger{\sum}}_{a} ~p_a \langle \psi | \rho_a | \psi \rangle  \big( \alpha + \beta \langle \psi | \rho_a | \psi \rangle\big) \notag \\
&=& 1 - \frac{\mathrm{dim}(\mathcal{H}^{\varovee N})}{d^{N-1} (d+1)} \mathlarger{\mathlarger{\sum}}_{a} ~p_a \langle \psi | \rho_a | \psi \rangle  \big( 1 + \langle \psi | \rho_a | \psi \rangle\big),
\ee 
as $\alpha=\beta=1/d(d+1)$. Now as we have an ensemble of Haar distributed pure states we replace the sum by an integral,
\be 
\int\D\mathcal{N} R_{\mathcal{N}}\big(\rho, \hat{\rho}^{\mathcal{N}}_B\big) &\leq & 1 -  \frac{\mathrm{dim}(\mathcal{H}^{\varovee N})}{d^{N-1} (d+1)}  \int  \D \psi_a \langle \psi | \psi_a \X \psi_a | \psi \rangle  \big( 1 + \langle \psi |  \psi_a \X \psi_a | \psi \rangle\big) \notag \\
& = & 1 -  \frac{\mathrm{dim}(\mathcal{H}^{\varovee N})}{d^{N-1} (d+1)}  \Tr \Big( \int  \D \psi_a | \psi_a \X \psi_a | \psi \X \psi|  + \notag \\
&~& ~~~~~~~~~~~~~~~~~~~~~~~~~~~~~~~~~~~~~~~~~~~~~\int  \D \psi_a |\psi_a \X \psi_a|^{\otimes 2} |\psi \X \psi|^{\otimes 2} \Big).
\ee 
Again, by Schur's lemma, the first integral is simply $\frac{\mathbb{1}}{d}$, and the second integral is the same as in Eqn. \eqref{eq:lem1:4}. It can be evaluated using Schur-Weyl duality as done earlier. We thus obtain 
\be 
\int\D\mathcal{N} R_{\mathcal{N}}\big(\rho, \hat{\rho}^{\mathcal{N}}_B\big) &\leq & 1 - \frac{\mathrm{dim}(\mathcal{H}^{\varovee N})}{d^{N-1} (d+1)} \Bigg( \frac1 d  + \frac{2}{d (d+1)} \Bigg) \notag \\
&=& 1 - \frac{(d+3)~\mathrm{dim}(\mathcal{H}^{\varovee N})}{d^{N} (d+1)^2},
\ee 
and as $\mathrm{dim}(\mathcal{H}^{\varovee N}) = \binom{N+d-1}{N}$ we obtain \eqref{eq:lem1}.
\epr 
The upper bound in \eqref{eq:lem1} is weak and  in the limit of large $N$ approaches the trivial bound for infidelity i.e. one. While it is clear that we cannot expect a $t$-design to derandomize this Bayesian procedure, we can ask if an approximate design might emerge. To this end, we numerically compare measurements defined through the Pauli group (1-design), a unitary 2-design, the Clifford group unitaries (unitary 3-design),  and Haar random bases for the case of a qubit in \cref{subsubsec:Clifford_vs_RMB}. 




We will now discuss the numerical implementation \cite[GitHub]{Quadeer_BME-RMB-PGM} of \cref{fig:algo1} in the following subsections using fidelity as the distance measure.  We consider three kinds of ensembles. One constituting of pure states sampled from the Haar distribution on $\mathcal{S}(\mathcal{H})$, another constituting of mixed states $\rho$ defined through complex matrices of the Ginibre ensemble of random matrices $\mathcal{G}(d)$, 
\be \label{eq:Ginibre-density}
\rho = \frac{G^\dagger G}{\Tr (G^\dagger G)}, ~~~~G \sim \mathcal{G}(d),
\ee  
and a third constituting of a uniform mix of states of all ranks for a given $d$. The ensembles consist of $L$ states and we assume a uniform prior over the sets. The Haar distributed pure states are generated by picking a fixed column of Haar distributed unitaries each time, and as $L$ approaches large values the uniform distribution over this finite set will approach the Haar measure on the set of pure states $\mathcal{S}(\mathcal{H})$. However, the situation is more complex for the case of mixed states as the Ginibre ensemble of random complex matrices induces a particular measure on the set of mixed states which is not uniform, and is called the Hilbert-Schmidt measure \cite{induced_measures_Zyczkowski_2001}. As $L$ approaches large values, a uniform distribution on the finite set of density matrices defined through \eqref{eq:Ginibre-density} should approach the Hilbert-Schmidt measure on the set of density matrices. The fact that a uniform measure over the set of density matrices concentrates around higher rank states for large $d$ implies that a typical density matrix would be highly mixed, and the Ginibre ensemble results in such density matrices by definition. For completeness, we consider a third kind of ensemble consisting of random quantum states of fixed ranks $1 \leq r \leq d$. For a given rank $r$, we pick as many uniformly distributed numbers between 0 and 1 and define a $r \times r$ diagonal matrix and rotate it using a $d\times r$ random unitary matrix (obtained through the QR decomposition of a Ginibre matrix). We simply consider a uniform mix of such states in our ensemble, and even though this will not result in a uniform measure over the set of density matrices, it suffices for our purposes. We discuss each of these cases further in the subsections below.

\subsubsection{Pure states}\label{subsec:1}
\begin{figure}[h!]
 	\includegraphics[width=0.9\textwidth]{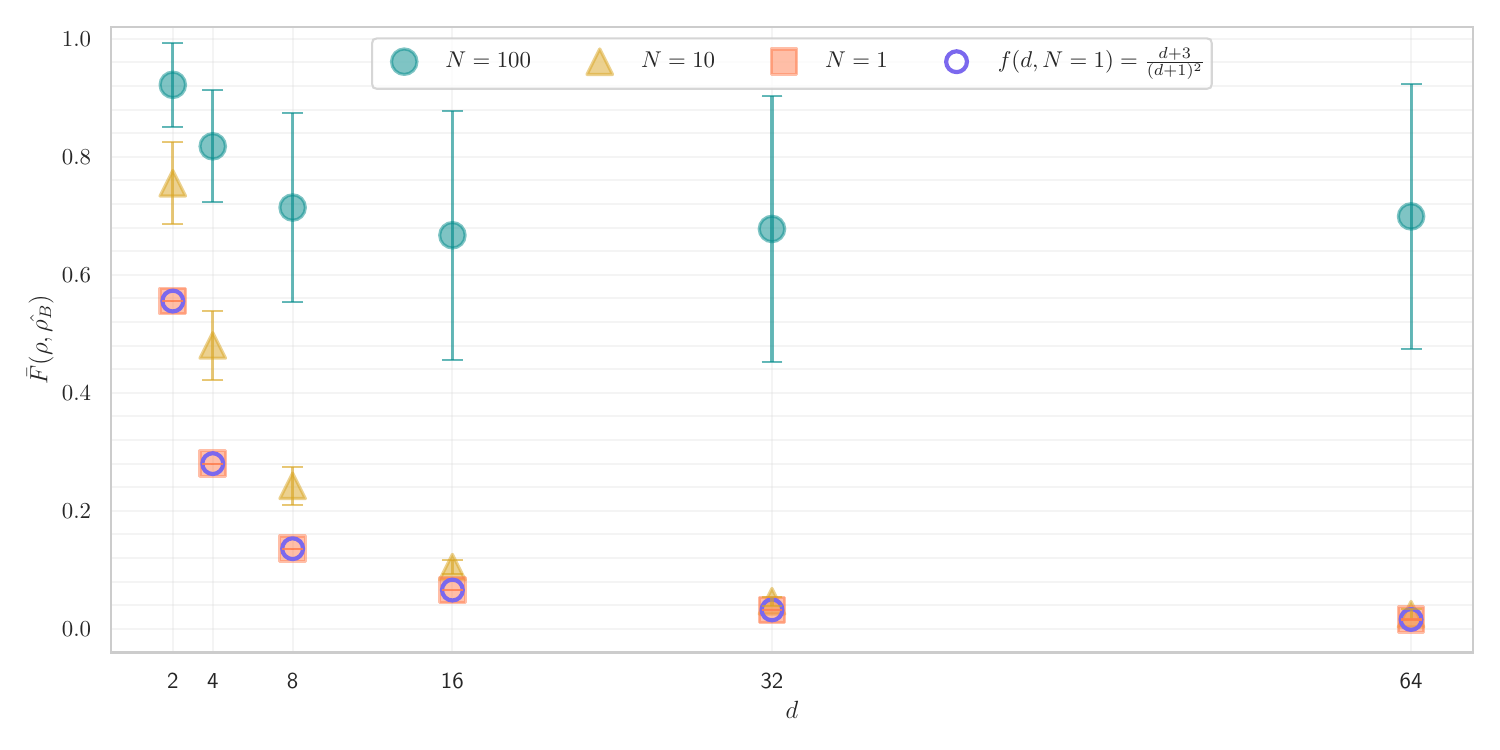} 
 	\caption{Average fidelity for Bayesian mean estimation $\bar{F}(\rho, \hat{\rho}_B)$ is shown as a function of the dimension $d$ for pure states with $N=1,10,100$. The analytically calculated value of $\bar{F}(\rho, \hat{\rho}_B)$ for $N=1$ is denoted by $f(d, N=1)$, \cref{lem:av-risk-pure-states}, and it matches the computed value. The relatively large error bars at $N=100$ are due to the underlying broad distributions of average risk itself as shown in \cref{fig:pure_distributions}, \cref{app:auxillary-plots}. The Haar random measurements are averaged over 1000 experiments each with $L=10^5$ states in the ensemble.}
 	\label{fig:pure}
 	\end{figure}

In \cref{fig:pure} we plot average fidelity defined via Eqn.~\eqref{eq:average-risk} as a function of dimension $d$ averaged over a set of $I=1000$ experiments. Each experiment constitutes of a sequence $\mathcal{N}$ of Haar random measurements. We obtain the average risk for a set of values of the number of measurements $N \in \{1, 10, 100\}$. For a single update corresponding to $N=1$, we derive an expression of the average risk which is independent of $\mathcal{U}$ in \cref{lem:av-risk-pure-states} and use it to validate the numerical algorithm, see \cref{proof-lem2} for proof.

\begin{lemma}\label{lem:av-risk-pure-states}
	The average risk of the Bayes estimator for a single measurement in a basis, defined through a unitary $\mathcal{U}$, for an ensemble of Haar distributed pure states $\mathcal{E}\sim\{\D\psi,|\psi\rangle\}$ in dimension $d$ is given by 
	\be \label{eq:lem:av-risk-pure-states}
	\int \D \psi~ R_{\mathcal{U}}\big(|\psi \X \psi |, \hat{\rho}_B\big) = 1 - \frac{d+3}{(d+1)^2},
	\ee 
	when risk is measured in terms of infidelity.
\end{lemma}
 The relatively large error bars are a consequence of the wide non-Gaussian distribution of the average risk itself which is in turn a consequence of the Haar distribution on $\mathbb{U}(d)$. We plot these distributions for a range of values of $I$ and $d$ in \cref{app:auxillary-plots}. 

\subsubsection{Mixed states}\label{subsec:2}
	\begin{figure}[h!]
 		\includegraphics[width=0.9\textwidth]{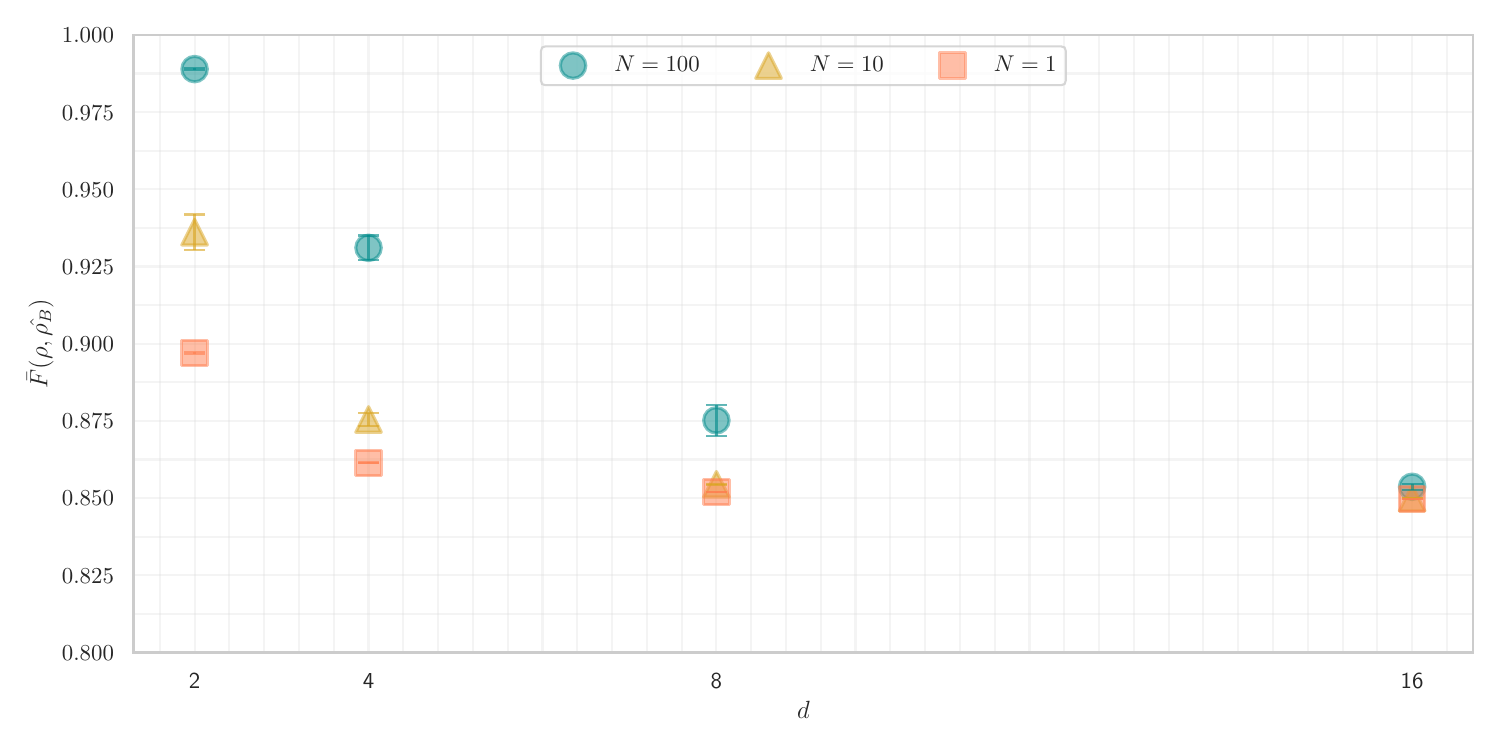}
 		\caption{Average fidelity for Bayesian mean estimation $\bar{F}(\rho, \hat{\rho}_B)$ is shown as a function of the dimension $d$ for mixed states with $N=1,10,100$. The small error bars at $N=100$ are due to the underlying distributions of average risk itself as shown in \cref{fig:mixed_distributions}, \cref{app:auxillary-plots}. The Haar random measurements are averaged over 1000 experiments each with $L=10^5$ states in the ensemble.}
 		\label{fig:mixed}
 	\end{figure}
 	
	\begin{figure}[h!]
 		\includegraphics[width=0.9\textwidth]{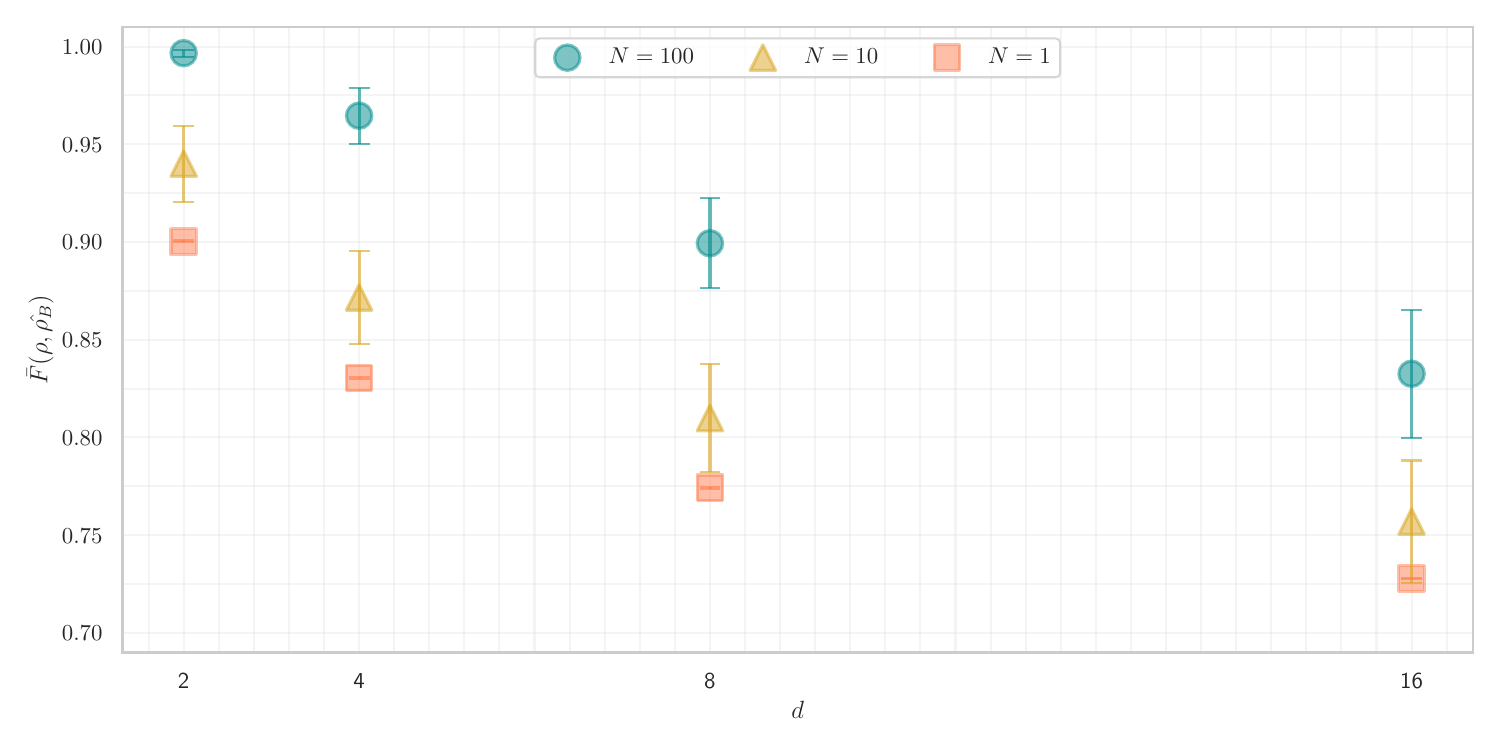}
 		\caption{Average fidelity for Bayesian mean estimation $\bar{F}(\rho, \hat{\rho}_B)$ is shown as a function of the dimension $d$ for a uniform mix of states of ranks $1 \leq r \leq d$ with $N=1,10,100$. The error bars at $N=100$ are due to the underlying distributions of average risk itself as shown in \cref{fig:mixed_distributions}, \cref{app:auxillary-plots}. The Haar random measurements are averaged over 1000 experiments each with $L=10^5$ states in the ensemble.}
 		\label{fig:rankdiff}
	\end{figure}
In \cref{fig:mixed} we plot the same quantities as in \cref{fig:pure} but for an ensemble of mixed states as defined by \eqref{eq:Ginibre-density}. Here we observe that the error bars are small---the averaging is over not very broad distributions. We have plotted the distributions for a set of values of $I$ and $d$ in \cref{fig:mixed_distributions}, \cref{app:auxillary-plots}. The fact that the Ginibre ensemble induces a Hilbert-Schmidt measure on the set of density matrices results in fairly mixed states on average \cite[Eqn. 3.16]{induced_measures_Zyczkowski_2001}, and gives us average fidelities that are not quite spread out when it comes to different sequences of Haar random measurements $\mathcal{N}$. Given that the states of the ensemble are mixed, and a convex sum of these will always lie within their convex hull, the Bayes estimator gives reasonably high fidelities on average at least for relatively small dimensions as we see in the plot. This is in contrast to what we have for the pure state ensemble, see for example the corresponding values for $d=16$. In order to see how Bayesian mean estimation does in general for states with different ranks we consider a uniform mix of states with ranks $1 \leq r \leq d$. In \cref{fig:rankdiff} we plot the average risk vs. the dimension $d$ averaged over a set of $I=1000$ experiments as before. The ensemble considered is a mix of $L$ states such that there are equal number of states of any given rank and thus the probability that any given state has a rank $r$ is $1/d$. While this distribution is still not uniform on $\mathcal{D}(\mathcal{H})$ it gives a reasonable spread in the qualitative properties of the states in the ensemble because of which we obtain a drop in the values of the average fidelity for corresponding dimensions in \cref{fig:mixed}. We plot the actual distribution of average fidelity in \cref{fig:rankdiff_distributions},  \cref{app:auxillary-plots} which give rise to the error bars and stem from the random sampling of the unitary matrices from $\mathbb{U}(d)$.
 
 \subsubsection{Clifford unitaries vs. Haar-random bases in $d=2$}\label{subsubsec:Clifford_vs_RMB}
 
 \begin{figure}[h!]
 	\includegraphics[width=0.9\textwidth]{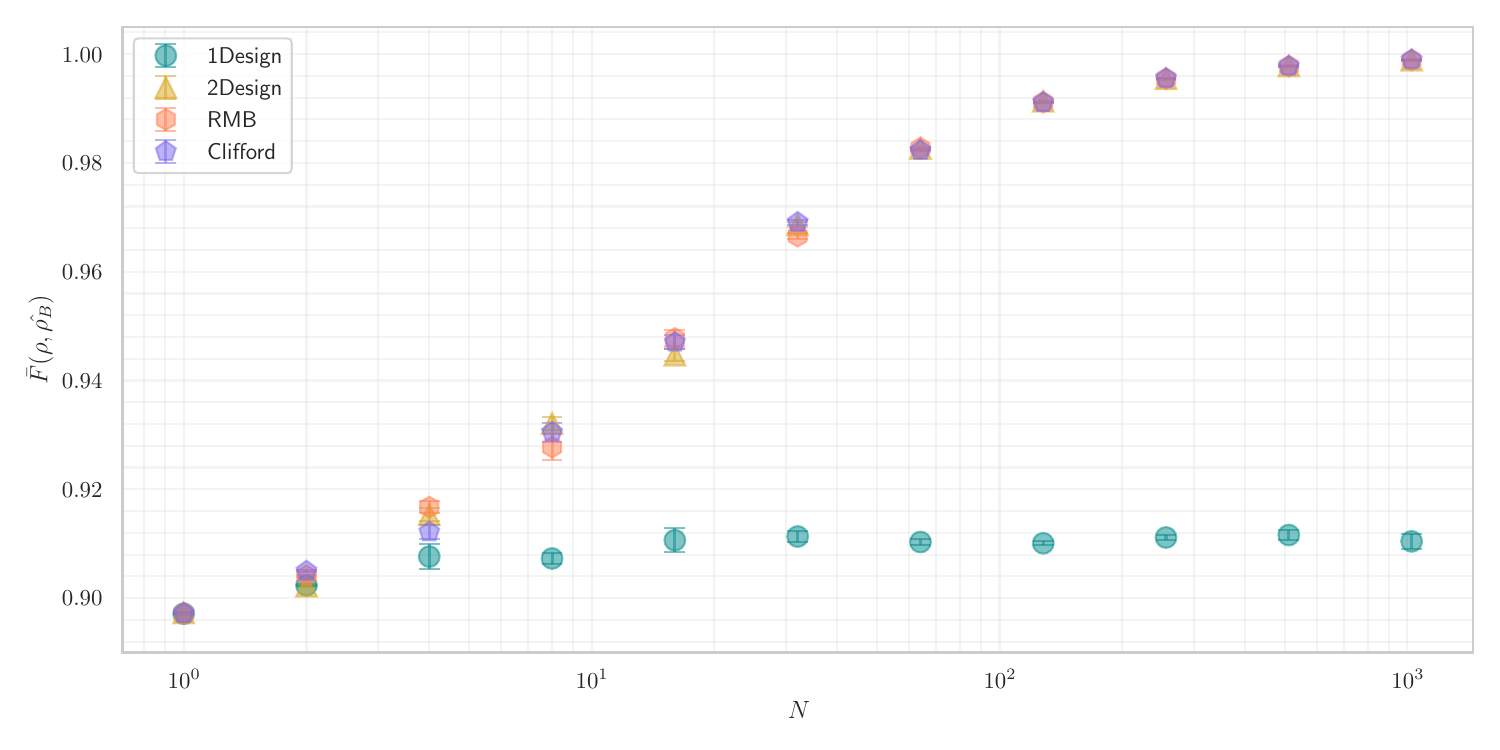}
 	\caption{Average fidelity for measurement bases defined through a unitary 1-design (1Design), a unitary 2-design (2Design), a unitary 3-design (Clifford), and Haar random unitaries (RMB) in $d=2$ is plotted as a function of the number of measurement shots $N$. There are $L=10^4$ states in the (uniform) ensemble of mixed states as in \cref{fig:mixed}.}
 	\label{fig:mixed_clifford_vs_rmb}
 \end{figure}
 
The issue of efficient implementation of Haar random measurements in \cref{subsec:2} still remains. But we also know that Haar averaging of a fractional polynomial will not result in a unitary $t$-design, see Eqn. \eqref{eq:Haar-av-risk-n-design-4}. This leaves room for obtaining bounds on risk in terms of $t$-designs in general. In \cref{fig:mixed_clifford_vs_rmb} we provide comparisons between the average fidelity for Haar random measurements and measurements defined through Clifford unitaries (which forms a $2$-design) in $d=2$ for an ensemble of mixed states. We find that unitary 2-designs closely approximate Haar random unitaries under average fidelity for ensembles of mixed qubit states while the Pauli group (that forms a 1-design) only gives a weak lower bound for average fidelity of Bayesian mean estimator. An extensive numerical analysis is needed to see if this would also hold for higher dimensions and we leave it to a future work.
 
\section{Bayesian mean estimation - Pretty-good measurements}\label{sec:BME-PGM}
Given an ensemble \eqref{eq:ensemble} of states, we define the quantity, 
\be 
\rho_{\mathrm{out}} = \sum_a \tilde{\rho}(a),
\ee 
and the corresponding square-root or pretty-good measurement (PGM) \cite{watrous_2018} via the POVM elements
\be \label{eq:PGM}
E_x = \rho_{\mathrm{out}}^{-1/2} ~\tilde{\rho}(x) ~\rho_{\mathrm{out}}^{-1/2}.
\ee 
In the following lemma we obtain the Bayes estimator for pretty good measurement (single-shot) and find that it has a simple expression due to the form of the POVM elements.
	
\begin{figure}[h!]
	\includegraphics[width=0.65\textwidth]{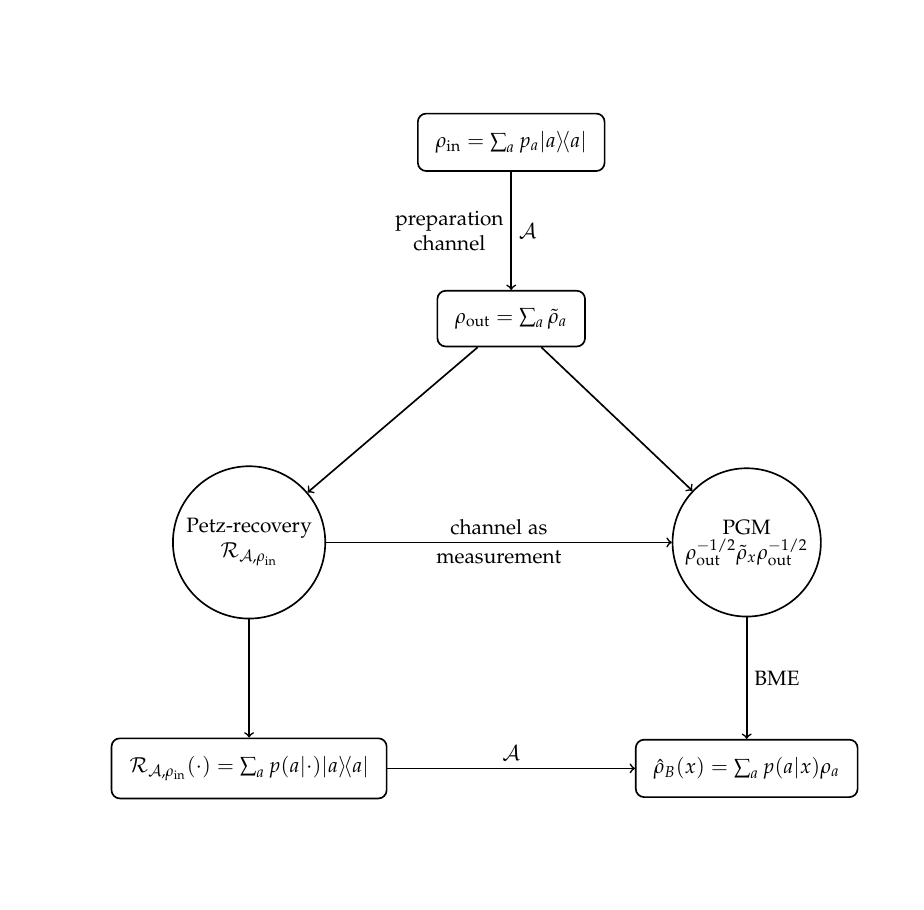}
	\caption{Bayesian mean estimation as Petz recovery.}
	\label{fig:1}
\end{figure}

\begin{lemma}[Covariance of output distribution for PGM under BME]\label{lem:cov-PGM}
The posterior distribution of states under pretty good measurement is simply the conditional distribution of the measurement outcomes with swapped indices, i.e.
\be \label{eq:lem:cov-PGM}
p(a|x) = \Tr (E_a\rho_x),
\ee 
where the index `x' corresponds to the measurement outcome, the index `a' corresponds to the state $\rho_a$, and the Bayes estimator is
\be\label{eq:BME-PGM}
	\hat{\rho}_B^{\mathrm{PGM}}(x) = \sum_a p(a|x)\rho_a.
\ee 
\end{lemma}

\bpr 
Eqns. \eqref{eq:Bayes-rule} and \eqref{eq:PGM} imply
\be 
p(a|x)
&=&  \frac{\Tr (E_x\rho_a)p_a}{p(x)} =  \frac{\Tr \big(\rho_{\mathrm{out}}^{-1/2} ~\tilde{\rho}(x) ~\rho_{\mathrm{out}}^{-1/2}\tilde{\rho}(a)\big)}{p(x)}.
\ee 
Now, $p(x)=\sum_a p(x|a)p_a = \sum_a \Tr \big(\rho_{\mathrm{out}}^{-1/2} ~\tilde{\rho}(x) ~\rho_{\mathrm{out}}^{-1/2}\tilde{\rho}(a)\big) = \Tr \big(\rho_{\mathrm{out}}^{-1/2} \rho_{\mathrm{out}} ~\rho_{\mathrm{out}}^{-1/2}\tilde{\rho}(x)\big)=p_x$. Thus, by \eqref{eq:PGM} we have
\be 
p(a|x)
&=& \Tr (\rho_{\mathrm{out}}^{-1/2} ~\tilde{\rho}(a)~\rho_{\mathrm{out}}^{-1/2} \rho_x),
\ee 
which implies \eqref{eq:lem:cov-PGM}. The expression for the Bayes estimator then simply follows by plugging \eqref{eq:lem:cov-PGM} into \eqref{eq:Bayes-est}. 
\epr 
In fact, the Petz recovery channel corresponding to pretty good measurement can be understood as Bayesian mean estimation by means of \cref{lem:petz-PGM}.
\begin{lemma} \label{lem:petz-PGM}
The Petz recovery map corresponding to a preparation channel $\mathcal{A}: |a\X a| \mapsto \rho_a$ is given by
	\be \label{eq:lem:petz-PGM}
	\mathcal{R}_{\mathcal{A},~\rho_{\mathrm{in}}}(\cdot) = \sum_a p(a|\cdot) |a \X a|,
	\ee 
	where $p(a|\cdot)$ is the posterior distribution under pretty good measurement.
\end{lemma}
\bpr 
Using the expression for the Petz recovery map as in Ref. \cite[Eqn. 16]{Petz-recovery_PGM_Barnum-Knill_2002} that corresponds to the preparation channel $\mathcal{A}$, and  assuming that $\rho_a = \sum_{b} \lambda_{ba} | v_{ba} \X v_{ba}|$, we obtain
\be\label{eq:petz-map}
\mathcal{R}_{\mathcal{A},~\rho_{\mathrm{in}}}(\cdot) &=& \sum_{a,~b} p_a ~\lambda_{ba} |a \X v_{ba}|~\rho_{\mathrm{out}}^{-1/2} (\cdot)\rho_{\mathrm{out}}^{-1/2} ~|v_{ba} \X a|\notag \\
&=& \sum_a p_a \sum_b \lambda_{ba} \Tr\bigg(\rho_{\mathrm{out}}^{-1/2} (\cdot)\rho_{\mathrm{out}}^{-1/2} |v_{ab} \X v_{ab}| \bigg) |a \X a|\notag \\
&=& \sum_a p_a \Tr\bigg(\rho_{\mathrm{out}}^{-1/2} (\cdot)\rho_{\mathrm{out}}^{-1/2} \rho_a \bigg) |a \X a| = \sum_a \Tr\bigg(\rho_{\mathrm{out}}^{-1/2} \tilde{\rho}(a)\rho_{\mathrm{out}}^{-1/2} (\cdot) \bigg) |a \X a|,
\ee 
and we get \eqref{eq:lem:petz-PGM} by \cref{lem:cov-PGM}.
\epr
Thus the fact that the output of the Petz recovery map for a preparation channel $\mathcal{A}$ is the mean of the input ensemble with respect to the posterior distribution implies the existence of a parallel between the Petz recovery channel and Bayesian mean estimation with PGM. The output of the Petz recovery process \eqref{eq:petz-map} and the corresponding Bayes estimator are related via the map $\mathcal{A}$. We illustrate this in \cref{fig:1}. 

\begin{figure}[h!]
 		\includegraphics[width=\textwidth]{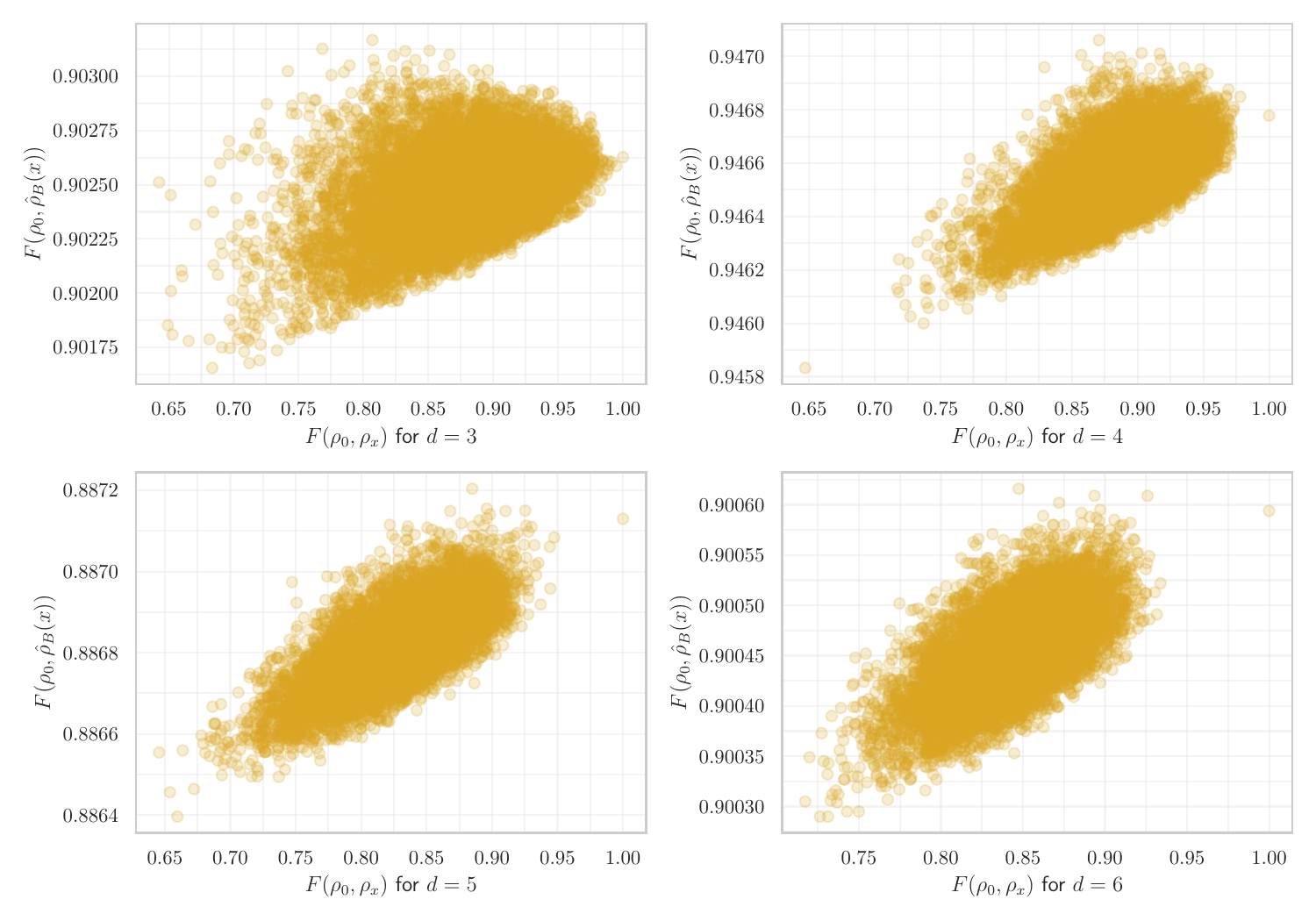}
 		\caption{$F(\rho_0, \rho_x)$ vs. $F(\rho_0, \hat{\rho}_B(x))$ for ensembles of mixed states defined through \eqref{eq:Ginibre-density} with $L=10^3$ states. The Bayes estimator is obtained using the pretty good measurement for the ensembles.}
 		\label{fig:PGM-1}
\end{figure}

In light of this connection we argue that Bayesian mean estimators are a natural choice when estimating states of an ensemble using pretty good measurement. We numerically show that there is a substantial improvement in fidelity between the given state and the estimate for a measurement outcome $x$ when using a Bayesian estimator $\hat{\rho}_B(x)$ as opposed to the naive estimator $\hat{\rho}(x)=\rho_x$. We quantify the respective fidelities by $F(\rho_0, \hat{\rho})$ for a fixed input state $\rho_0$. We then plot this quantity for $\hat{\rho}(x) = \rho_x$ against the corresponding value for $\hat{\rho}(x) = \hat{\rho}_B(x)$ for fixed dimensions $d$ in \cref{fig:PGM-1}. However, the nature of the Ginibre ensemble (of mixed) states along with the fact that the Bayes estimator lies in the convex hull of the ensemble could effect this outcome. 

\section{Discussion \& Outlook}\label{sec:discuss-outlook}
In this work we studied Bayesian mean estimation under two different measurement settings. One constituting of Haar random measurements under a sequential updates algorithm and the other involving PGM for obtaining single-shot Bayesian mean estimates. We showed that Bayesian mean estimation under PGM is essentially Petz recovery by showing that the output distribution of PGM is covariant under a Bayesian update. For the case of Haar random measurements under a sequential updates algorithm, we discovered expressions involving Haar averages of rational functions over the unitary group, Eqn. \eqref{eq:Haar-av-risk-n-design-4}. While analytical closed form expressions of such averages are still unknown \cite{kostenberger2021weingarten}, it is clear that such averages cannot be obtained using unitary $t$-designs. We showed a comparison between measurements defined through a unitary $1$-design (Pauli group), a unitary $2$-design, a unitary 3-design (Clifford group), and Haar random unitaries for an ensemble of mixed qubit states. We found that a unitary $2$-design gives a good approximation to the Haar random bases while the Pauli group only gives a weak lower bound. However, an exhaustive numerical study for higher dimensions is needed to shed light on whether an approximate $2$-design might work, and we leave this direction of enquiry to a future work.

\begin{acknowledgements}
I thank Marco Tomamichel for a useful discussion and feedback on the manuscript, Chris Ferrie for providing a reader's lens, and Geet Rakala for help with the plots. This work was supported by the start-up grant of the Nanyang Assistant Professorship awarded to Nelly H.Y. Ng of Nanyang Technological University, Singapore.
\end{acknowledgements}

\bibliography{references}

\begin{appendix}
\section{Proof of \cref{lem:av-risk-pure-states}}\label{proof-lem2}
\bpr 
Using Eqns. \eqref{eq:risk} and \eqref{eq:p(x|i)}, and the expression for fidelity when one of the states is pure, we obtain
\be \label{eq:lem:av-risk-pure-states:1}
R_{\mathcal{U}}\big(|\psi \X \psi |, \hat{\rho}_B\big) = 1 - \mathlarger{\sum}_x~ \Tr\big(\mathcal{U}|x \X x| \mathcal{U}^\dagger |\psi \X \psi |\big) \langle \psi | \hat{\rho}_B(x) |\psi \rangle.
\ee 
Substituting the expression for the Bayes estimator using Eqn. \eqref{eq:Bayes-est} above along with \eqref{eq:Bayes-rule}, we have 
\be \label{eq:lem:av-risk-pure-states:2}
R_{\mathcal{U}}\big(|\psi \X \psi |, \hat{\rho}_B\big) &=& 1 - \mathlarger{\sum}_{x,~a}~ p_{\mathcal{U}}(a|x)\Tr\big(\mathcal{U}|x \X x| \mathcal{U}^\dagger |\psi \X \psi |\big) \langle \psi | \rho_a |\psi \rangle \notag \\
&=& 1 - \mathlarger{\mathlarger{\mathlarger{\sum}}}_{x,~a}~ \frac{p_{\mathcal{U}}(x|a)p_a}{p_{\mathcal{U}}(x)}\Tr\big(\mathcal{U}|x \X x| \mathcal{U}^\dagger |\psi \X \psi |\big) \langle \psi | \rho_a |\psi \rangle.
\ee 
Now, Eqns. \eqref{eq:Bayes-rule} and \eqref{eq:p(x|i)} together imply $p_{\mathcal{U}}(x)=\frac{1}{d}$, thus 
\be \label{eq:lem:av-risk-pure-states:3}
R_{\mathcal{U}}\big(|\psi \X \psi |, \hat{\rho}_B\big) &=& 1 - d\mathlarger{\mathlarger{\mathlarger{\sum}}}_{x,~a}~ p_{\mathcal{U}}(x|a)p_a \Tr\big(\mathcal{U}|x \X x| \mathcal{U}^\dagger |\psi \X \psi |\big) \langle \psi | \rho_a |\psi \rangle \notag \\
&=& 1 - d\mathlarger{\mathlarger{\mathlarger{\sum}}}_{x}~ \Tr\big(\mathcal{U}|x \X x| \mathcal{U}^\dagger |\psi \X \psi |\big)\sum_a~ p_{\mathcal{U}}(x|a)p_a  \langle \psi | \rho_a |\psi \rangle.
\ee 
Let us focus on the second sum 
\be \label{eq:lem:av-risk-pure-states:4}
\sum_a~ p_{\mathcal{U}}(x|a)p_a  \langle \psi | \rho_a |\psi \rangle &=& \sum_a~ p_a  \Tr\big(\mathcal{U}|x \X x| \mathcal{U}^\dagger \rho_a \big) \langle \psi | \rho_a |\psi \rangle \notag \\
&=& \sum_a~ p_a \langle x_{\mathcal{U}}| \rho_a |x_{\mathcal{U}} \rangle \langle \psi | \rho_a |\psi \rangle,
\ee 
where we set $|x_{\mathcal{U}}\rangle = \mathcal{U}^\dagger |x\rangle$. The ensemble of pure states reduces to the following integral over the Haar ensemble in the limit of a large $L$,
\be \label{eq:lem:av-risk-pure-states:5}
\lim_{L \rightarrow \infty}~ \mathlarger{\sum}_a~ p_a \langle x_{\mathcal{U}}| \rho_a |x_{\mathcal{U}} \rangle \langle \psi | \rho_a |\psi \rangle &=& \int \D \phi ~\langle x_{\mathcal{U}}| \phi \X \phi |x_{\mathcal{U}} \rangle \langle \psi | \phi \X \phi |\psi \rangle \notag \\
&=& ~\Tr \int \D \phi | \phi \X \phi |^{\otimes 2}~ |\psi \X \psi| \otimes | x_{\mathcal{U}} \X x_{\mathcal{U}}|.
\ee 
We can do the integral above using Schur-Weyl duality to obtain 
\be \label{eq:lem:av-risk-pure-states:6}
\int \D \phi ~\langle x_{\mathcal{U}}| \phi \X \phi |x_{\mathcal{U}} \rangle \langle \psi | \phi \X \phi |\psi \rangle &=& ~\Tr  \big(\alpha \mathbb{1} + \beta \mathbb{W}\big) |\psi \X \psi| \otimes | x_{\mathcal{U}} \X x_{\mathcal{U}}| \notag \\
&=& \frac{1}{d(d+1)} \big(1 + \Tr|\psi \X \psi| x_{\mathcal{U}} \X x_{\mathcal{U}}|\big),
\ee 
where $\alpha=\beta=1/d(d+1)$. Plugging \eqref{eq:lem:av-risk-pure-states:6} in \eqref{eq:lem:av-risk-pure-states:3} and switching back to $|x_{\mathcal{U}}\rangle = \mathcal{U}^\dagger |x\rangle$ we obtain 
\be 
\int \D \psi~ R_{\mathcal{U}}\big(|\psi \X \psi |, \hat{\rho}_B\big) 
&=& 1 - \frac{1}{d+1}\Bigg( 1 + \mathlarger{\mathlarger{\mathlarger{\sum}}}_{x}~ \Tr~ \int \D \psi~ |\psi \X \psi|^{\otimes 2} ~\mathcal{U}^{\otimes 2}|x \X x|^{\otimes 2} (\mathcal{U}^\dagger)^{\otimes 2}  \Bigg). 
\ee 
Again by Schur-Weyl duality we have 
\be 
\int \D \psi~ R_{\mathcal{U}}\big(|\psi \X \psi |, \hat{\rho}_B\big) 
&=& 1 - \frac{1}{d+1}\Bigg( 1 + \mathlarger{\mathlarger{\mathlarger{\sum}}}_{x}~ \Tr~ \big(\alpha \mathbb{1} + \beta \mathbb{W}\big) ~\mathcal{U}^{\otimes 2}|x \X x|^{\otimes 2} (\mathcal{U}^\dagger)^{\otimes 2}  \Bigg) \notag \\
&=&  1 - \frac{1}{d+1}\Bigg( 1 + \mathlarger{\mathlarger{\mathlarger{\sum}}}_{x}~ \frac{2}{d(d+1)}\Bigg),
\ee 
as $\alpha=\beta=1/d(d+1)$, and thus the above gives \eqref{eq:lem:av-risk-pure-states}.
\epr 

\newpage
\section{Auxillary plots}\label{app:auxillary-plots}
\begin{center}
	\begin{figure}[h!]
	\includegraphics[width=1.0\textwidth]{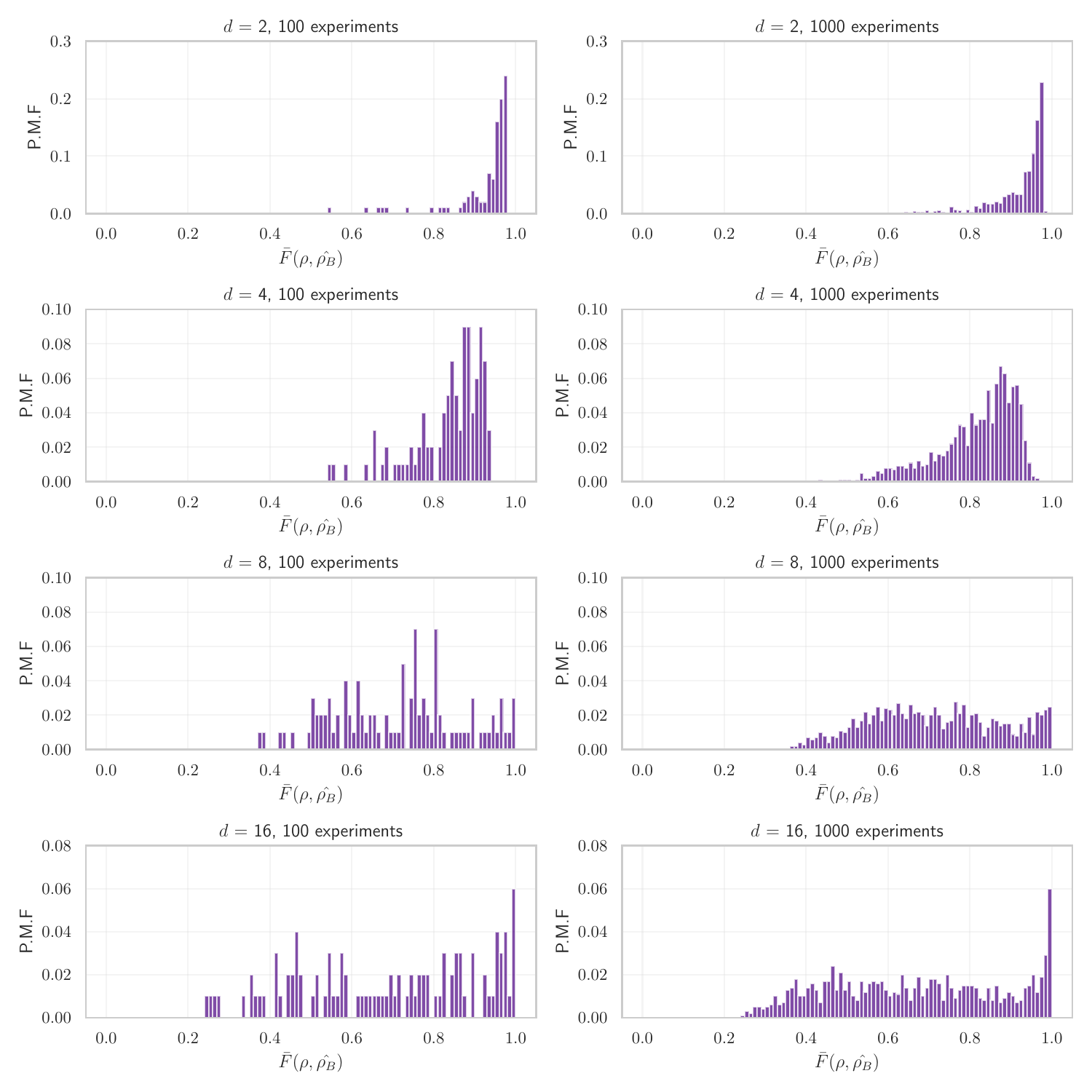}
	\caption{Distribution of average risk for $N=100$ and ensemble of pure states. Increasing the number of experiments does not reduce the variance in average risk of \cref{fig:pure}.}
	\label{fig:pure_distributions}
	\end{figure}
\end{center}
\begin{center}	
	\begin{figure}[h!]
	\includegraphics[width=1.0\textwidth]{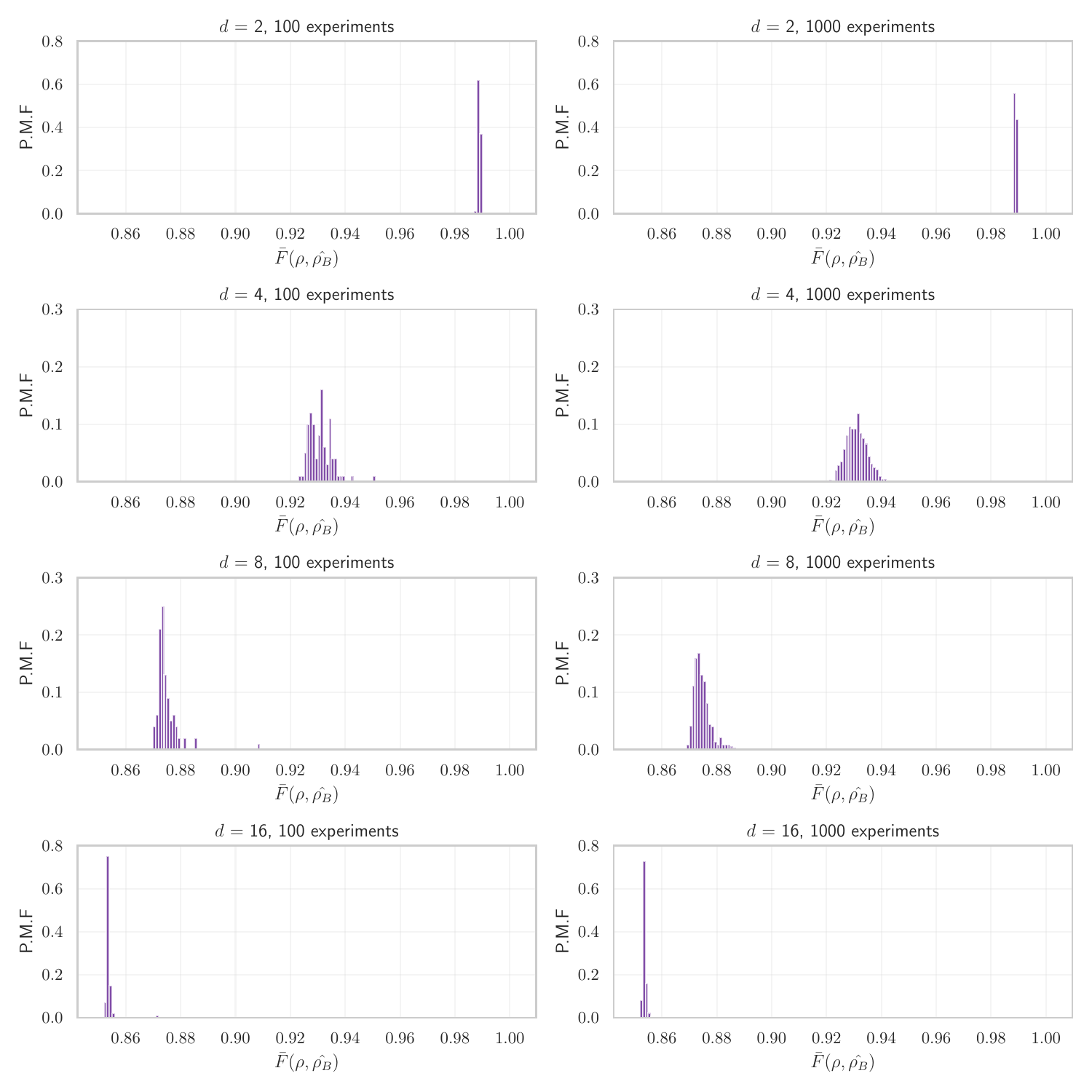}
	\caption{Distributions of average risk for $N=100$ and ensemble of mixed states. Increasing the number of experiments does not reduce the variance in average risk of \cref{fig:mixed}.}
	\label{fig:mixed_distributions}
	\end{figure}
\end{center}
\begin{center}
	\begin{figure}[h!]
	\includegraphics[width=1.0\textwidth]{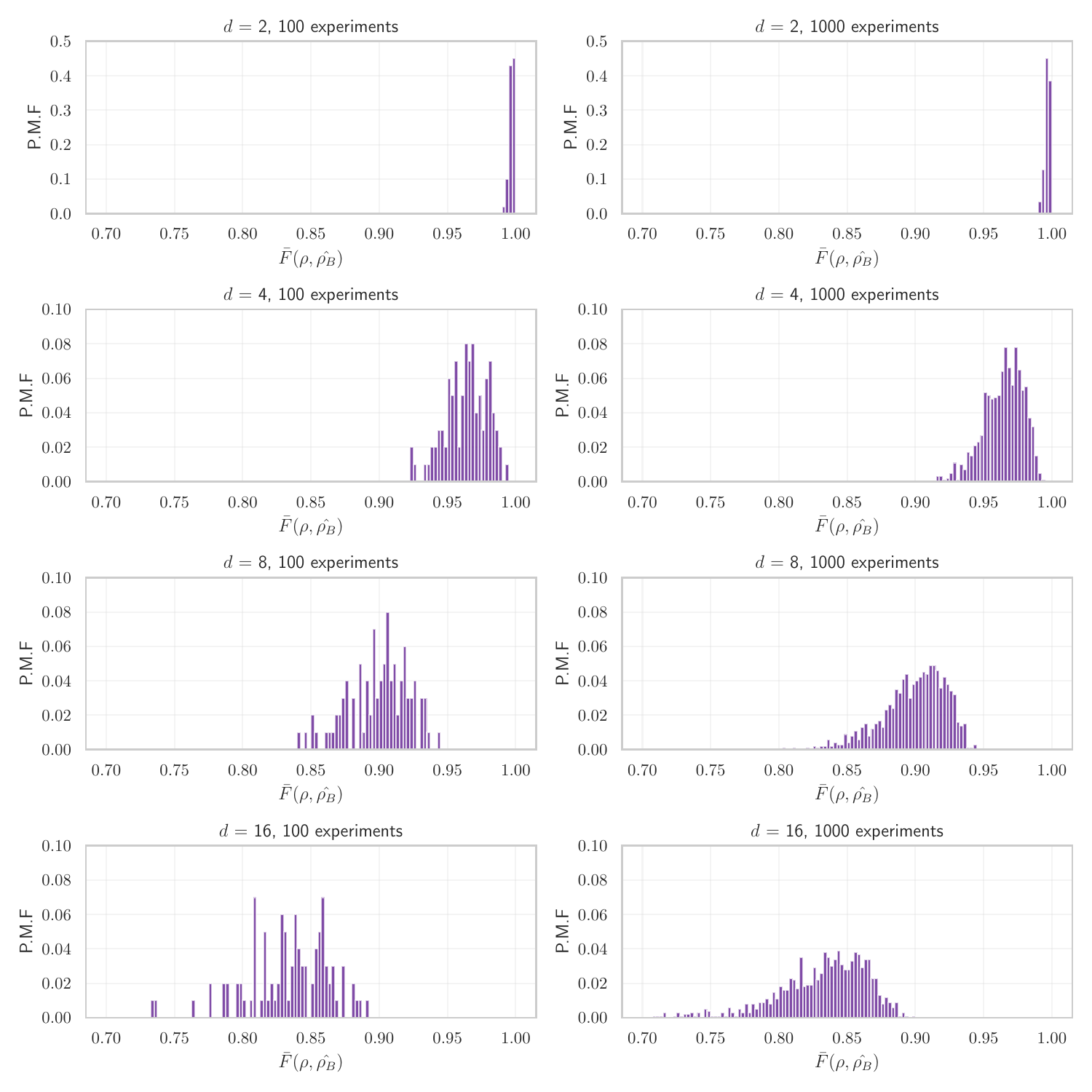}
	\caption{Distributions of average risk for $N=100$ and ensemble of states with ranks $1 \leq r \leq d$. Increasing the number of experiments does not reduce the variance in average risk of \cref{fig:rankdiff}.}
	\label{fig:rankdiff_distributions}
	\end{figure}
\end{center}

\end{appendix}

\end{document}